\documentclass[preprint]{nature}
\usepackage{graphicx}
\usepackage{subfigure}
\usepackage{dcolumn}
\usepackage{bm}
\usepackage[mathlines]{lineno}
\usepackage{array}
\usepackage{braket}
\usepackage{diagbox}
\usepackage{amsmath}
\usepackage{appendix}
\usepackage{multirow}
\usepackage{bm,color,bbm}

\usepackage{float}
\newcolumntype{.}{D{.}{.}{-1}}
\usepackage{amsopn}
\usepackage[colorlinks, linkcolor=red, anchorcolor=black, urlcolor=red, citecolor=blue]{hyperref}
\usepackage{epstopdf}

\usepackage{booktabs}
\usepackage{enumerate}
\usepackage{siunitx}
\usepackage{upgreek}
\usepackage{amssymb}
\usepackage{caption}
\usepackage{mathrsfs}

\begin{document}
\title{SHAPE: a sparse Hamiltonian approach for prescribing eigenstates in metamaterials}
\author
{Yutao Chen$^{1,2,3,\dagger}$, Tao Xiao$^{1,\dagger}$, Jing Li$^{1,\dagger}$, Aodong Li$^{1}$, Biye Xie$^{1,\ast}$ }
\maketitle

\begin{affiliations}
    \item School of Science and Engineering, The Chinese University of Hong Kong (Shenzhen),  Shenzhen, Guangdong, 518172, China
	\item Shanghai Research Center for Quantum Science and CAS Center for Excellence in Quantum Information and Quantum Physics, University of Science and Technology of China, Shanghai 201315, China
	\item Hefei National Laboratory, University of Science and Technology of China, Hefei 230088, China
	\\ 
	$\dagger$ These authors contributed equally.

	$^\ast$ Corresponding author.

	Email: xiebiye@cuhk.edu.cn
\end{affiliations}
	
\begin{abstract}
Metamaterials, which store, guide and emit waves through their eigenstates, have enabled the development of devices with diverse functionalities.
Yet  the spatial profiles of conventional metamaterials' eigenstates are typically determined by their pre-determined structures (Hamiltonians).
Here we develop SHAPE, a sparse Hamiltonian approach that enables the realization of prescribed, spectrally isolated eigenstates in physically realizable metamaterials.
The resulting Hamiltonian geometries obey a boundary law, with the required nonlocality determined solely by single-contact boundary sites (\(\lvert E_{\mathrm{LR}}\rvert = \lvert \partial\Omega \rvert\)), independent of the state interior or lattice size. 
The Hamiltonians are further spectrally reshaped to open a protective gap around the target eigenstate.
We validate our approach in discrete lattices and  acoustic metamaterials.
We  provide a database of over 100 target states and a customizable design program, demonstrating the generality and extensibility of our method.
Although demonstrated in acoustics, the principle can be applied to thermal, electrical-circuit and photonic metamaterials.
\end{abstract}
	
\maketitle
	
\begin{figure}[!htb]
		\centering
		\includegraphics[width=1\linewidth]{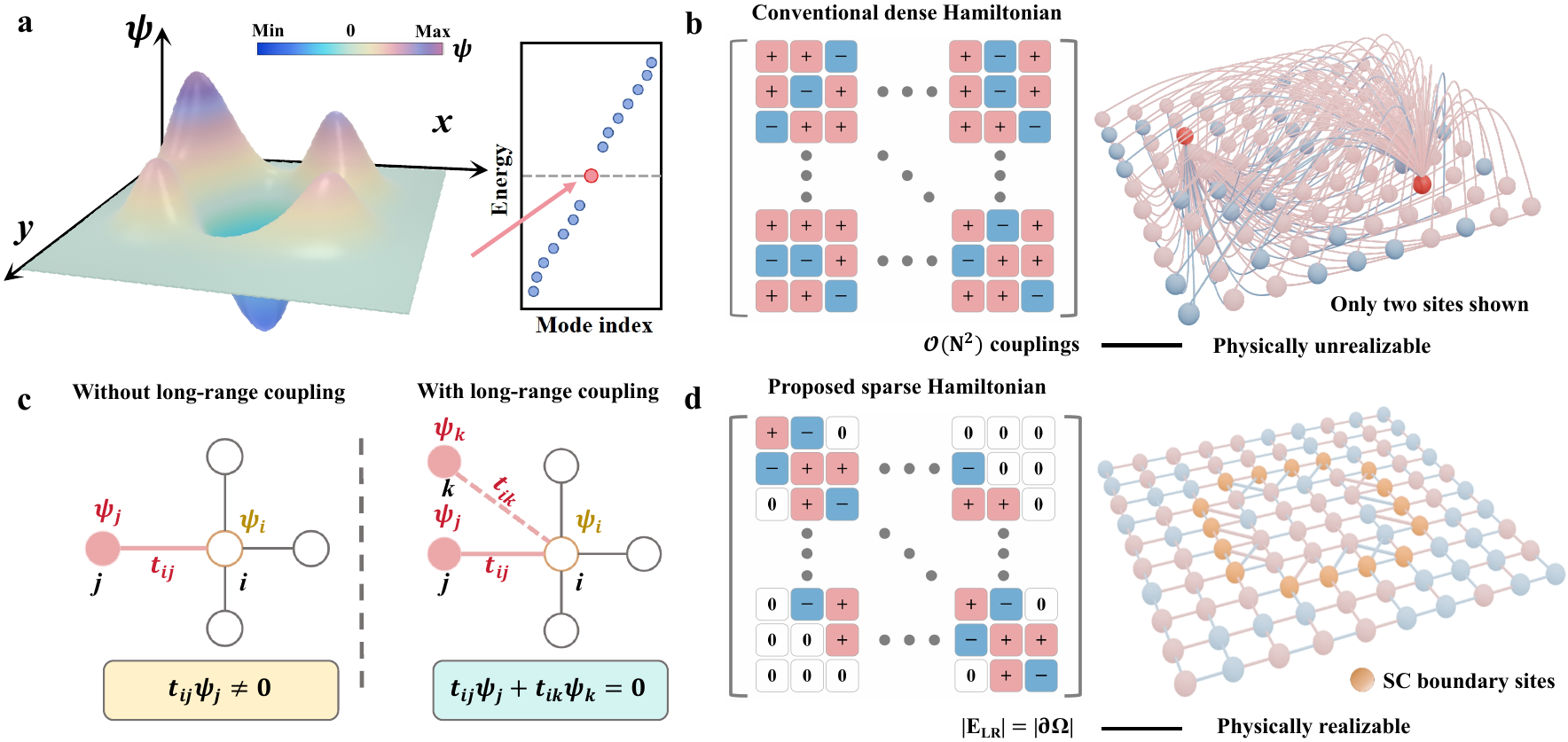}
		\caption{\textbf{Boundary-law-guided realization of metamaterials with prescribed eigenstates.}
\textbf{a},  Eigenstate with prescribed spatial profile and spectral isolation.
\textbf{b}, Dense Hamiltonian via direct spectral construction, with \(\mathcal{O}(N^2)\) couplings, making physical implementation impractical. For clarity, only the all-to-all connections of two sites are shown. 
\textbf{c}, An SC boundary site is a zero-amplitude site connected to exactly one nonzero-amplitude site, and only at such sites does the local destructive-interference condition require a long-range coupling.
\textbf{d}, Sparse Hamiltonian of the proposed metamaterials, consisting predominantly of nearest-neighbor couplings, with additional long-range couplings introduced only at the SC boundary sites highlighted in orange. The total number of long-range couplings satisfies \(\lvert E_{\mathrm{LR}}\rvert = \lvert \partial\Omega \rvert\).}
		\label{fig1}
	\end{figure}

Metamaterials offer a versatile platform for controlling waves through engineered structures and interactions. 
Among the physical mechanisms underlying this control, eigenstates play a central role in wave localization, confinement and energy transport.
Yet in most known settings, the spatial  profiles of these eigenstates are inherited from the host, such as  Anderson localization\cite{anderson1958,schwartz2007}, defect modes in photonic crystals\cite{yablonovitch1991donor,akahane2003highq,painter1999}, 
bound states in the continuum\cite{hsu2013trapped,hsu2016bic}, topological
boundary states\cite{hasan2010ti,lu2014topophotonics,yang2015topoacoustics,xue2019kagome,susstrunk2015mechanical,ni2019hoti},  and compact localized states\cite{leykam2018flatband,Maimaiti2021}.
Numerical methods follow the same direction\cite{Sha2025}, from a given structure to its spectrum and modes.
This raises a basic question: can a field distribution be specified  and then realized as a gap-isolated eigenstate in a physically implementable metamaterial (Fig.~\ref{fig1}a)?
Achieving this would reverse the conventional Hamiltonian-to-wavefunction relation, elevating the spatial profile of an eigenstate from a consequence of the host to an independently controllable physical degree of freedom.

The challenges lie in precisely prescribing eigenstate spatial profiles, ensuring spectral isolation and maintaining the physical realizability of the Hamiltonian.
Algebraic constructions based on spectral decomposition can prescribe an eigenstate and its eigenvalue by completing an orthogonal basis\cite{Chu2005}, but generally yield dense Hamiltonians with nearly all-to-all couplings (Fig.~\ref{fig1}b), which are difficult to implement in photonic\cite{hafezi2013imaging}, acoustic\cite{liu2000lrsm,lu2009phononic,zhang2018topologicalsound,ma2019topophases} and electronic\cite{kiczynski2022engineering} lattices. 
Inverse Hamiltonian methods have been developed to tailor dispersion relations and band structures\cite{lucas2023tailoring,kazemi2023drawing} or construct parent Hamiltonians for prescribed many-body eigenstates\cite{qi2019determining,chertkov2018computational}, but generally do not provide precise control over the spatial profiles of eigenstates.
Wavefront-shaping metasurfaces and acoustic holograms can generate prescribed fields\cite{yu2011metasurface,li2017controlling,melde2016holograms}, yet these are typically externally driven responses rather than intrinsic, spectrally isolated eigenstates.

Here we address these challenges by introducing a general framework SHAPE:  a sparse Hamiltonian approach for prescribing spectrally isolated eigenstates in  physically realizable metamaterials.
The  target  prescribed eigenstate is first embedded through site-wise local destructive-interference conditions, with long-range couplings introduced only at the single-contact (SC) boundary sites.
This leads to a boundary law: the required nonlocal couplings are determined solely by the SC boundary sites of the wavefunction, independent of its interior or lattice size, yielding sparse, physically realizable Hamiltonians with \(\lvert E_{\mathrm{LR}}\rvert = \lvert \partial\Omega \rvert\).
The Hamiltonian is subsequently Hermitianized, followed by spectral reshaping to open a protective gap around the target eigenstate.
We provide a comprehensive theoretical foundation for this approach and validate it through four examples in discrete lattices, full-wave continuum models and acoustic metamaterials. 
We also provide a database of over 100 target states and a customizable program for generating new ones. 
The principle can also be applied to thermal, electrical-circuit and photonic metamaterials described by tight-binding models.
We believe this study opens a new, practical route to controlling prescribed eigenstates in metamaterials.

\paragraph{Results.}
\subsection{Sparse Hamiltonian construction} 

Consider a general two-dimensional real-valued tight-binding model defined on a lattice with $N$ sites,
\begin{equation}
\bm{H}=\sum_i \epsilon_i\, c_i^{\dagger}c_i
+\sum_{i}\sum_{j\in\mathcal{N}_i} t_{ij}\, c_i^{\dagger}c_j
+\sum_{i}\sum_{j\in\mathcal{N}^{\mathrm{LR}}_i} t'_{ij}\, c_i^{\dagger}c_j
 ,
\label{eq:ham}
\end{equation}
where $\epsilon_i$ is the on-site energy, $\mathcal{N}_i$ and $\mathcal{N}^{\mathrm{LR}}_i$ denote the sets of nearest neighbors and long-range neighbors of site $i$, and $t_{ij}$, $t'_{ij}$ are the corresponding couplings. 
We construct a sparse Hamiltonian that embeds the prescribed target state as follows. 
For a real-valued state $\bm{\psi}_0$ at energy $\lambda_0$, the eigenvalue equation $\bm{H}\bm{\psi}_0=\lambda_0\bm{\psi}_0$ decomposes into one scalar condition at each lattice site,
\begin{equation}
(\epsilon_i-\lambda_0)\,\psi_0(i)
+\sum_{j\in\mathcal{N}_i} t_{ij}\,\psi_0(j)
+\sum_{j\in\mathcal{N}^{\mathrm{LR}}_i} t'_{ij}\,\psi_0(j)=0,\ i=1,2,\cdots, N.
\label{eq:sitecond}
\end{equation}
Each equation represents a local destructive-interference condition at the corresponding site, requiring all amplitudes coupling into site 
$i$ to cancel at the target energy, so the state is pinned by interference imposed bond by bond.
We start from a Hermitian Hamiltonian containing only on-site terms and nearest-neighbor couplings, and enforce the destructive-interference condition site by site. 
Among these sites, there exists a special class of SC boundary sites, defined as $\partial\Omega = \{i \in K : b_i = 1\}$, where $K = Z(\bm{\psi}_0) = \{i : \psi_0(i) = 0\}$ and $b_i := \bigl|\{\,j \in \mathcal{N}_i : \psi_0(j) \neq 0\,\}\bigr|$. 
Note that the SC boundary sites defined here are zero-amplitude sites located along the periphery of the given wavefunction and connected to only one nonzero-amplitude nearest neighbor, without introducing any real physical boundary.

At the SC boundary sites, the destructive-interference condition contains only a single non-zero product between a nearest-neighbor coupling and $\psi_0(j)$. Satisfying the local balance therefore requires an additional long-range coupling to a non-zero-amplitude site, as illustrated in Fig.~\ref{fig1}c.
The remaining nearest-neighbor coupling cannot vanish, because this would locally disconnect the lattice and create a physical boundary, rather than embed the target eigenstate in the full Hamiltonian.
At all other sites, the condition can be satisfied by adjusting only a single nearest-neighbor coupling or an on-site term. 
This observation leads to the boundary law stating that a prescribed multi-site wavefunction $\bm{\psi}_0$ on a two-dimensional lattice can be embedded as an exact eigenstate, with
\begin{equation}
|E_{\mathrm{LR}}| = |\partial\Omega|,
\end{equation}
where $|E_{\mathrm{LR}}|$ is the number of long-range couplings and $|\partial\Omega|$ is the number of SC boundary sites.
See Supplementary Notes 1 and 2 for more detailed construction and proof.

This boundary law guarantees sparsity of the resulting Hamiltonian and dictates the number and positions of added long-range couplings.
The construction according to the boundary law avoids the severe overconstraint of direct sparse-matrix synthesis\cite{Chu2005} by enforcing sparsity through local destructive-interference conditions rather than by explicitly constraining a large number of matrix elements to zero, thereby preserving sufficient degrees of freedom for subsequent Hamiltonian manipulation.
Moreover, the boundary law also offers guidance by linking the required long-range couplings to the geometry of the target wavefunction. 
By choosing profiles that eliminate the SC boundary site condition, the number of long-range couplings can be reduced and, in some cases, removed entirely, enabling a purely nearest-neighbor realization.

The sparse Hamiltonian obtained by embedding the target state can be directly Hermitianized. 
The embedding procedure described above enforces the destructive-interference condition row by row, but does not by itself guarantee that the resulting matrix is Hermitian. 
We find, however, that whenever the interference condition at a site $i$ is satisfied by adjusting a coupling $H_{ij}$, whether nearest-neighbor or long-range, the reciprocal matrix element can directly be set to $H_{ji}=H_{ij}^{*}$ without disrupting the destructive-interference condition at site $j$ (see Supplementary Note 3). 
The Hamiltonian can therefore be maintained Hermitian after each site-wise adjustment simply by imposing the conjugate reciprocal coupling. 
This operation introduces no additional long-range couplings, but only enforces reciprocity among the couplings already present. 
Thus, the sparse Hamiltonian embedding the prescribed target state can always be constructed as a Hermitian matrix without altering either the target eigenstate or its sparsity structure.

The target eigenstate can be isolated as an in-gap mode through spectral reshaping of the sparse Hermitian Hamiltonian within a linear subspace.
A uniform energy shift, sets the target eigenvalue to zero, so that $\bm{\psi}_0\in\ker\bm{H}$, without altering the eigenvectors or the sparsity pattern. 
We then construct a linear subspace $V$ containing $\bm{H}$ (see Supplementary Note 4), in which every matrix shares the same zero-eigenvalue target state and sparsity pattern as $\bm{H}$.
Hence, $\bm{H}$ can be expressed as a linear combination of a complete set of basis matrices $\{\bm{B}_k\}$ of $V$,
\begin{equation}
\bm{H}(\bm{\theta})=\sum_{k=1}^{m}\theta_k\bm{B}_k, \qquad \theta_k\in\mathbb{R}. \end{equation}
This establishes a one-to-one correspondence between the admissible Hamiltonians in $V$ and the unconstrained parameters $\bm{\theta}=(\theta_1,\cdots,\theta_m)$, providing a general representation of these special Hamiltonians in terms of unconstrained parameters.

Consequently, $\bm{\theta}$ can be varied freely to reshape the remaining eigenvalue spectrum without altering either the zero-energy target state or the prescribed sparsity pattern of $\bm{H}(\bm{\theta})$, thereby spectrally isolating $\bm{\psi}_0$.
The length of $\bm{\theta}$, i.e., the dimension of the subspace $V$, is given by
\begin{equation}
\dim V = |E_{\mathrm{NN}}| + |E_{\mathrm{LR}} | +|\Omega^{e}|
\end{equation}
where $|E_{\mathrm{NN}}|, |E_{\mathrm{LR}}|  $ denote the numbers of  nearest-neighbor couplings and long-range couplings, respectively, and $\Omega^{e}=\left\{ i \in K : b_i=0\right\} $  denotes the set of exterior sites.
Every additional nearest-neighbor bond, long-range bond, or exterior site enlarges $V$, and with it the freedom to reshape the spectrum around $\bm{\psi}_0$.  
See Supplementary Note 5 for the corresponding proof.  
For a fixed coupling number, the dimensional relation provides an algebraic interpretation of why the achievable spectral isolation typically becomes more restricted as the prescribed state extends over a larger region.
As the target state occupies more lattice sites, fewer exterior sites remain, reducing the dimension of $V$ and therefore the freedom available to displace the surrounding eigenstates.
This result is consistent with the usual physical intuition that a single-mode eigenstate with a larger spatial support is generally less amenable to isolation by a large spectral gap. 
The basis $\left\{\bm{B}_k\right\}$ can be chosen in multiple ways, and a direct construction is provided in Supplementary Note 6.

Based on the parameterized representation of $\bm{H}(\bm{\theta})$, 
we define the normalized spectral gap around the target state as
\begin{equation}
g(\bm{\theta})
=
\frac{
|\lambda|_{(2)}\!\left(\bm{H}(\bm{\theta})\right)
}{
|\lambda|_{(N)}\!\left(\bm{H}(\bm{\theta})\right)
},
\label{eq:gap}
\end{equation}
where $|\lambda|_{(2)}$ and $|\lambda|_{(N)}	$ denote, 
respectively, the second-smallest and largest absolute eigenvalues of 
$\bm{H}(\bm{\theta})$. 
The former gives the nearest spectral separation from the prescribed 
zero-energy state $\bm{\psi}_0$, whereas the latter sets the overall spectral 
scale of the Hamiltonian.
The normalized gap function $g(\bm{\theta})$ is shown to possess two important properties.
First, the case in which the target eigenstate is single-mode is open, dense, and of full measure in parameter space, once a single-mode eigenstate exists (see Supplementary Note 7).
Second, for nonzero $\bm{\theta}$, $g(\bm{\theta})$ is locally Lipschitz 
continuous and differentiable almost everywhere with respect to the Hamiltonian parameters (see Supplementary Note 8). 
These regularity properties establish a well-defined local spectral landscape and enable gradient-guided deterministic reshaping of the surrounding spectrum within ${V}$, thereby enhancing the spectral isolation of the prescribed state.


This completes our construction of a sparse Hamiltonian using the prescribed wavefunction as a blueprint, with Hermiticity enforced and the target state spectrally isolated.
Note that although the framework and theory above have been formulated for real-valued Hamiltonians and eigenstates, they remain applicable to complex-valued field distributions when complex-valued Hamiltonians are allowed (see Supplementary Note 9).


\subsection{Numerics based on tight-binding model} 
\begin{figure}[p]
		\centering
		\includegraphics[width=0.9\linewidth]{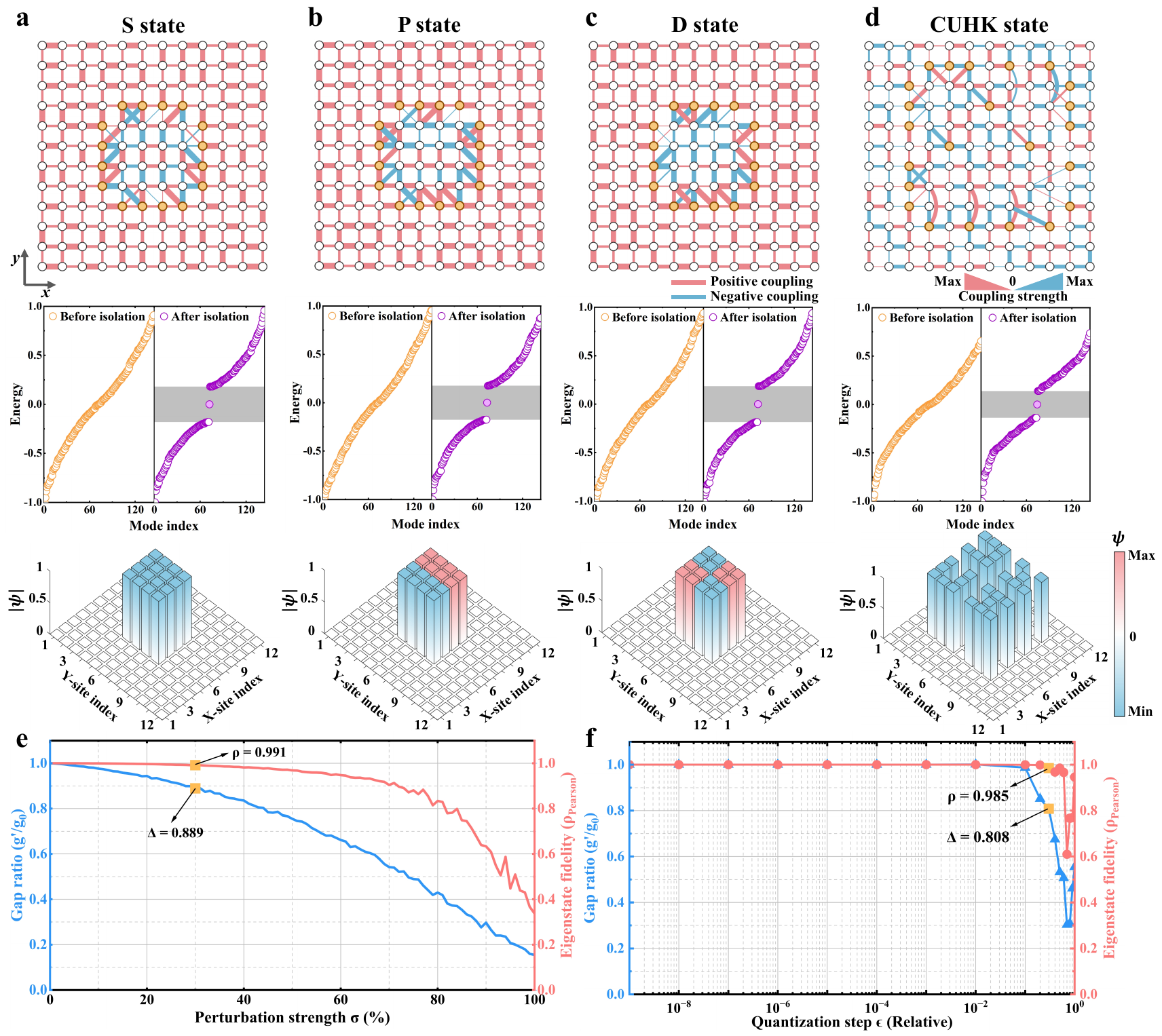}
		\caption{\textbf{Prescribed eigenstates, spectral isolation and robustness in tight-binding lattices.}
			\textbf{a--d,} Realization of the prescribed orbital-like $S$, $P$, $D$, and CUHK-shaped eigenstates, respectively.
			Top, sparse structures of the corresponding Hamiltonians, with orange-filled circles marking SC boundary sites. Red and blue lines indicate positive and negative couplings, respectively, with line widths proportional to coupling strengths.
			Middle, the eigenvalues before (left) and after (right) gap opening, respectively. The target eigenvalue is fixed at zero energy and appears as the isolated magenta circle within the grey shaded gap.
			Bottom, target eigenstates, with bar height showing $|\psi|$, red and blue indicating positive and negative amplitudes.
			\textbf{e, f,} Robustness of target states against random Hamiltonian perturbations (\textbf{e}) and quantization with relative step $\epsilon$ (\textbf{f}). Blue and red curves indicate the normalized spectral gap and Pearson correlation with the original eigenstate, respectively.}
\label{fig2}
	\end{figure}

We implement a range of prescribed eigenstates on a $12\times 12 $ two-dimensional square lattice,  from three orbital-like profiles ($S$,
$P$ and $D$ states) to a highly asymmetric, letter-shaped state.
Within the target region, the $S$ state is defined to have a uniform sign, the $P$ state is divided into two opposite-sign domains by a nodal line, and the $D$ state exhibits a quadrupolar sign pattern.
Realizing these states demonstrates the ability of the proposed framework to embed spatially confined eigenstates with precisely prescribed field distributions.
The letter-shaped state forms the four-letter pattern ``CUHK'', further validating the capability for a strongly asymmetric and spatially structured target profile.
The corresponding Hamiltonians, eigenvalue spectra, and target-state field distributions are shown in Fig.~\ref{fig2}.

In all four cases the latteice stays connected, with no isolated pieces or internal boundaries (Fig.~\ref{fig2}a--d, top), so each target is an eigenstate of the whole lattice.
Every long-range bond starts at one of the marked SC boundary sites, and the numbers of long-range bonds (16, 16, 16 and 23 for the $S$, $P$, $D$ and CUHK-shaped states) equal the numbers of SC boundary sites (Supplementary Table 1).
The interior sites of target states, irrespective of their number or the complexity of their field distributions, require no additional nonlocal couplings.

In each spectrum the target eigenvalue sits alone inside a gap (Fig.~\ref{fig2}a--d, middle).
We quantify this isolation using the normalized gap defined in Eq.~\eqref{eq:gap}. 
Before spectral reshaping, these target states are embedded within a dense manifold of spectator modes, with the largest spectral gap reaching only $g=0.25\%$.
Moving within the subspace $V$ raises $g$ to $18.03\%$, $17.63\%$, $18.51\%$ and $13.84\%$ for the four modes, corresponding to more than a fiftyfold enhancement.
The in-gap eigenvectors (Fig.~\ref{fig2}a--d, bottom) agree with the prescribed profiles to numerical precision (squared deviation below $10^{-29}$; Supplementary Table 1) and vanish outside the target regions.

A fabricated sample can never reproduce the ideal matrix exactly, so we test two kinds of error on all four targets (Fig.~\ref{fig2}e,f).
For the random perturbations, independent errors of prescribed magnitude are added to every coupling and on-site term, with 30 realizations generated for each perturbation strength and each mode. 
For the discretization test, all Hamiltonian entries are quantized using a series of step sizes. 
We record the ramaining gap, relative to its unperturbed value, and the Pearson correlation between the perturbed eigenvector and the target, averaged over the four targets and, for random errors, over all realizations.
At a random perturbation strength of $30\%$, the target states are recovered with aPearson correlation of $0.991$ and retain $88.9\%$ of the original gap. 
Likewise, undercoarse discretization with a step size of $30\%$, they are recovered with a Pearsoncorrelation of $0.985$ and retain $80.8\%$ of the original gap. 
The gap is what protectsthe target. 
By Weyl's inequality each eigenvalue moves by at most $\/\delta H\/$, so thetarget stays isolated as long as $\|\delta H\|$ is less than half the gap; by the Davis--Kahan theorem its eigenvector rotates by an angle of order $\|\delta H\|/\Delta$, where $\Delta$ is the gap.

Beyond these examples, to demonstrate the broader potential and generality of the approach, we have realized more than 100 single-mode eigenstates, including  letters, numerals and geometric shapes, canonical mathematical wavefunctions, and characteristic shapes from classical condensed-matter and quantum physics.
For each, we provide the spatial distribution, the spectrum and the coupling structure of the Hamiltonian (see Supplementary Note 10). 
Building on the principle, we also develop and release a general program that designs single-mode eigenstates of prescribed shape on demand from user input; the program is openly
and freely available, with documentation and usage instructions provided (see Supplementary Note 11).

The boundary law can also be used to control the number and the locations of long-range couplings. 
For some special eigenstates, the number of SC boundary sites can even be reduced to zero, allowing the target state to be realized entirely with nearest-neighbor couplings.
This is the same local canellation that underlies compact localized states in flat-band lattices\cite{Maimaiti2021}; here it yields a single gap-isolated state rather than a degenerate band.
By definition, an SC boundary site is a zero-amplitude site with exactly one nearest neighbor of nonzero amplitude. 
Therefore, by breaking this local condition, for example, through an appropriate choice of the wavefunction contour or by modifying the lattice geometry, the SC boundary sites can be shifted to different locations or eliminated.
We verify this for several specially structured eigenstates with zero SC boundary sites, all of which are realized entirely through nearest-neighbor couplings (see Supplementary Note 12). 

The approach is not fundamentally limited to real-valued eigenstates and applies equally to complex-valued eigenstates when complex, phase-carrying couplings are permitted. We further validate this generality through several complex-valued eigenstates, as provided in Supplementary Note 13.

\subsection{Design in continuum acoustic models}
We next ask whether the construction survives in a continuum wave system.
This is not guaranteed, because the acoustic metamaterial, unlike the discrete model, is governed by the continuum Helmholtz equation and inevitably involves frequency-dependent couplings, finite-sized resonators and residual higher-order interactions. 
We bridge these descriptions using an acoustic cavity--tube architecture, in which each vertical cavity provides a local resonant degree of freedom and the connecting tubes implement the inter-site hopping amplitudes~\cite{wang2023zclassified}.

To simplify the acoustic implementation, we constrained all on-site energies to a common value (zero after a shift); all but one are exactly zero, and the last is driven close to zero (Supplementary Note 14).
All sites can then be identical cavities, and only the tubes need to be calibrated.
The couplings were then rounded to multiples of 0.1.
With the constraint and the rounding, the realized target modes deviate from the prescribed states by at most $5.6\times10^{-4}$ in squared norm, and the numbers of long-range couplings become 14, 14, 14 and 32 for the $S$, $P$, $D$ and CUHK samples (Supplementary Note 14 and Table 1).

With the on-site terms fixed, the mapping between the off-diagonal Hamiltonian elements and the connecting-tube geometries is calibrated using a two-cavity building block (Fig.~\ref{fig4}a). 
Coupling hybridizes the local cavity resonance into even and odd modes, whose frequency splitting determines the effective hopping strength. 
The tube attachment position and length primarily control the coupling magnitude and sign, respectively. 
The resulting design map in Fig.~\ref{fig4}b spans continuously tunable positive and negative couplings, providing the dictionary required to translate the sparse Hamiltonians into manufacturable acoustic structures. 

In this way we designed acoustic versions of the four targets, using the constrained and rounded Hamiltonians.
Finite-element simulations of the $12\times12$ structures show, in each case, a single eigenstate inside a spectral window near the design frequency (Fig.~\ref{fig4}c).
The simulated eigenfrequencies are $2{,}321.7$, $2{,}321.1$, $2{,}313.1$ and $2{,}328.2~\mathrm{Hz}$ for the $S$, $P$, $D$ and CUHK states, respectively. 
These results show that both the prescribed field profiles and their protective gaps persist beyond the tight-binding approximation in the full continuum acoustic system. 
The full-wave simulation settings and field-extraction procedure are described in Methods.

The corresponding simulated pressure fields are shown in Fig.~\ref{fig4}d--g.
The outer panels show the pressure in the full structure; the central maps show the pressure sampled in each cavity, which plays the role of the tight-binding amplitude.
The projected fields reproduce the prescribed target profiles and their intended sign structures, with the $S$ state showing an almost uniform phase over the central region, the $P$ state exhibiting a single nodal division, the $D$ state displaying the expected quadrupolar sign pattern, and the CUHK state retaining its extended, asymmetric letter-shaped distribution.
The agreement between these projected fields and the prescribed target states demonstrates that the tight-binding design can be faithfully carried into a continuum acoustic system, establishing its physical feasibility and providing a concrete basis for realization in a real acoustic platform.

The continuum spectra differ quantitatively from the discrete tight-binding predictions. 
The acoustic structures provide an effective realization of the target matrices, with local reactive loading, frequency-dependent couplings and residual scattering beyond the discrete model.
These effects shift the effective on-site frequencies and couplings slightly and by different amounts in each structure, which contributes to the spread of the four eigenfrequencies (up to $15~\mathrm{Hz}$).
Nevertheless, the essential features required for the present construction are preserved, including the spectral isolation of the target state, the fidelity of its pressure profile to the prescribed eigenvector, and the sparsity of the Hamiltonian at the level of physical connectivity.

\subsection{Experimental observations in acoustic metamaterials}
To experimentally validate the target eigenstates, we fabricated the designed acoustic metamaterials using stereolithographic 3D printing, as depicted in the upper panels of Fig.~\ref{fig5}a--d. 
Samples \#1-4 realize the compact orbital-like $S$, $P$  and $D$  states and asymmetric CUHK-shaped state, respectively.
Each sample consists of a $12\times12$ array of vertical acoustic cavities connected by tubes whose geometries were chosen from the calibrated tight-binding-to-continuum mapping (see Methods). 

\begin{figure}[!htb]
		\centering
		\includegraphics[width=1\linewidth]{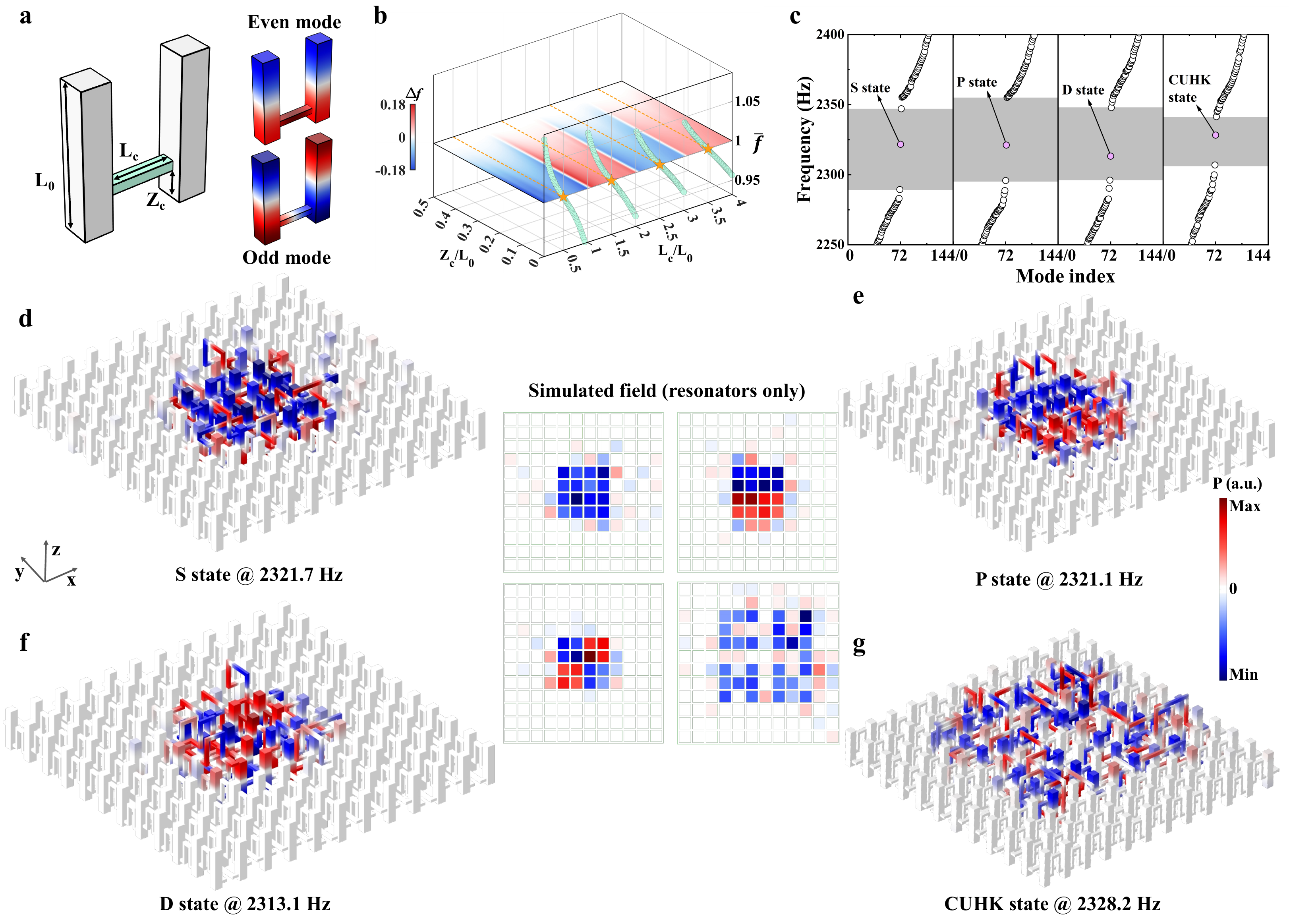}
		\caption{\textbf{Mapping sparse Hamiltonians to continuum acoustic eigenstates.}
			\textbf{a,} Two-cavity unit used to calibrate the coupling against the tube geometry. The right panel shows the even and odd modes.
			\textbf{b,} The map used for designing acoustic structures. Normalized eigenfrequency splitting $\Delta f$ and mean frequency $\bar{f}$ as functions of $L_c/L_0$ and $Z_c/L_0$. Orange stars and dashed lines mark representative geometries satisfying $\bar{f}=1$.
			\textbf{c,} Full-wave eigenfrequency spectra of the acoustic metamaterials designed for the $S$, $P$, $D$ and CUHK-shaped states. Target states (magenta circles) are marked within the shaded spectral windows. 
			\textbf{d--g,} Simulated pressure fields of the $S$, $P$, $D$ and CUHK-shaped target states, respectively, at the corresponding eigenfrequencies marked in \textbf{c}. The outer three-dimensional plots show the full pressure fields in the acoustic structures, and the central maps show the corresponding fields sampled at the resonator sites. Red and blue denote positive and negative pressure amplitudes, respectively.}
		\label{fig4}
	\end{figure}

\begin{figure}[!htb]
		\centering
		\includegraphics[width=1\linewidth]{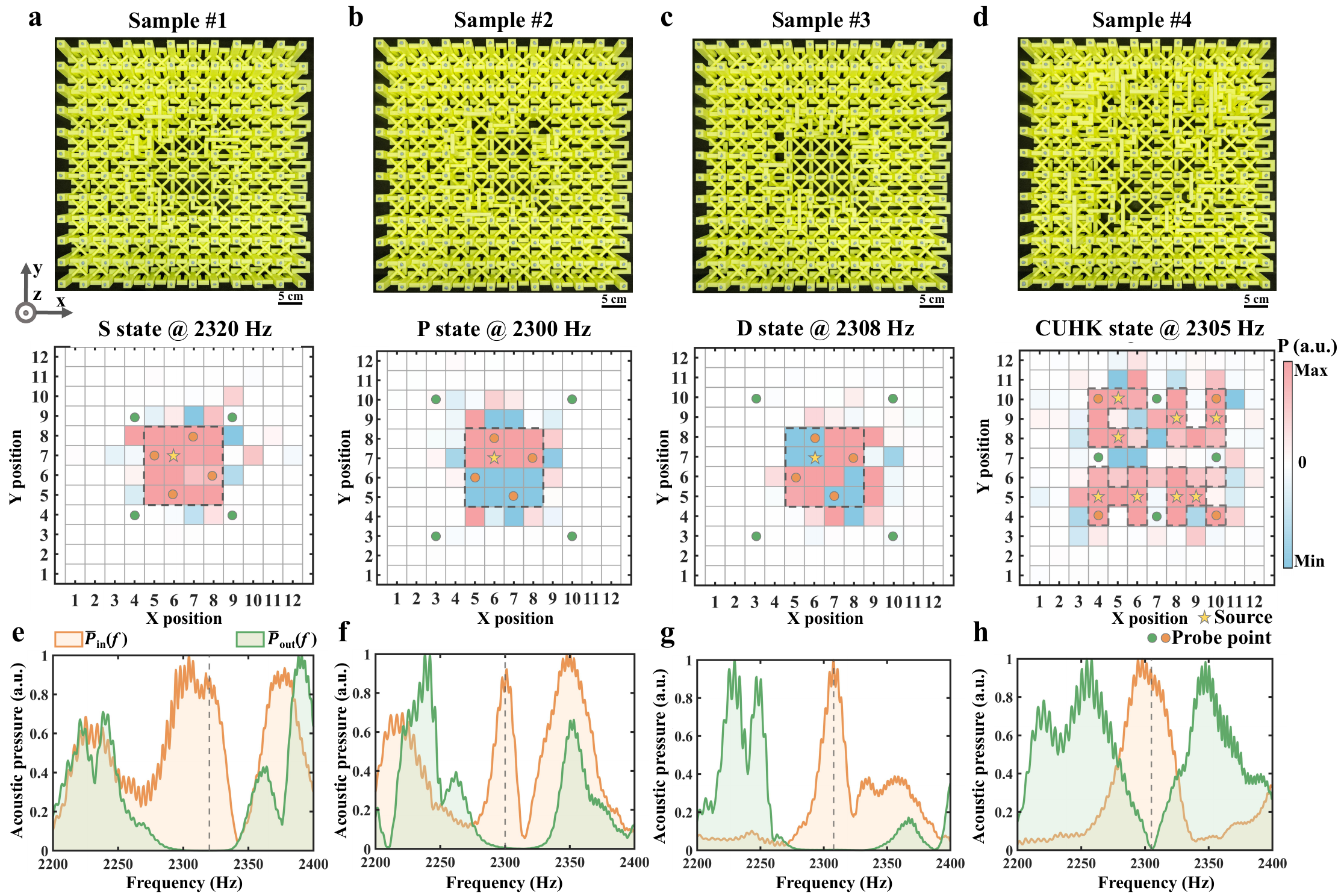}
	    \caption{\textbf{Experimental observation and in-gap spectroscopy of prescribed acoustic states.}
	    	\textbf{a--d,} Experimental realizations of the prescribed orbital-like $S$, $P$, $D$ and CUHK-shaped states, respectively.
	    	Upper panels show fabricated acoustic metamaterials with $12\times12$ cavities. 
	    	Lower panels show the measured acoustic pressure distributions at the labelled resonance frequencies. Color represents the signed acoustic pressure, with red and blue denoting positive and negative values, respectively, and white indicating zero pressure. Yellow stars mark the source positions, while orange and green circles denote probes inside and outside the regions of the prescribed states, respectively. The CUHK-shaped state is reconstructed by coherently combining measurements from eight source positions.
	    	\textbf{e--h,} Measured spectral responses of the corresponding samples. Orange and green curves show $\overline{P}_{\mathrm{in}}(f)$ and $\overline{P}_{\mathrm{out}}(f)$, averaged over probe points inside and outside the prescribed state regions, respectively. Vertical dashed lines mark the frequencies of the measured field maps in \textbf{a--d}.}
	\label{fig5}
	\end{figure}

To visualize the eigenstates, we excited each sample from selected source positions and measured the phase-resolved acoustic pressure field over the resonator array (see Methods).
For the compact orbital-like $S$, $P$ and $D$ states, a single source within the characteristic region is sufficient to excite the target resonance.
The CUHK state is spread over four separate letters, so we excited it in turn from eight non-nodal sites, two per letter, to improve coverage and signal-to-noise ratio.
All eight source sites are selected from the same-phase lobe of the target state.
For each excitation, the measured complex field is referenced to the simultaneously recorded complex pressure in the source cavity and normalized in amplitude.
The eight normalized complex fields are then coherently averaged to reconstruct the CUHK-shaped field (see Methods).
The measured maps (Fig.~\ref{fig5}a--d, bottom; $2{,}320$, $2{,}300$, $2{,}308$ and $2{,}305~\mathrm{Hz}$ for the $S$, $P$, $D$ and CUHK) reproduce the features predicted by the simulations in Fig.~\ref{fig4}, apart from an overall sign set by the reference phase.

We then analyzed the spectral isolation of these experimentally observed states using spatially resolved spectroscopy.
For each sample, two sets of probe points are selected.
The orange points lie inside the characteristic modal region, corresponding to the central target block for the $S$, $P$ and $D$ states, and the letter-shaped support for the CUHK state.
The green points lie outside this region and sample the surrounding lattice.
From the measured complex pressure $p_i(f)$, we construct two normalized spectral responses,
\begin{equation}
{P}_{\mathrm{in(out)}}(f)
=
\frac{1}{N_{\mathrm{in(out)}}}
\sum_{i\in \mathcal{R}_{\mathrm{in(out)}}}
\left|p_i(f)\right|,
\qquad
\overline{P}_{\mathrm{in(out)}}(f)
=\frac{P_{\mathrm{in(out)}}(f)}
{\max_{f'\in\mathcal{F}}P_{\mathrm{in(out)}}(f')},
\end{equation}
where $\mathcal R_{\mathrm{in}}$ and $\mathcal R_{\mathrm{out}}$ denote the sets of selected probe sites inside and outside the prescribed support, $N_{\mathrm{in}}$ and $N_{\mathrm{out}}$ are their respective sizes, and $\mathcal F$ is the measured frequency range.

As illustrated in Fig.~\ref{fig5}e--h, the orange curves show $\overline{P}_{\mathrm{in}}(f)$, while the green curves show $\overline{P}_{\mathrm{out}}(f)$.
In all four samples $\overline{P}_{\mathrm{in}}$ has a clear peak near the target frequency, whereas $\overline{P}_{\mathrm{out}}$ has none (Fig.~\ref{fig5}e--h).
This indicates that the target mode is spectrally separated from the extended resonances of the lattice.
The measured frequencies are $1.7--23.2~\mathrm{Hz}$ below the simulated ones, which we attribute to fabrication tolerances and viscothermal losses.

\paragraph{Discussion.}
We have shown that a field pattern chosen in advance can be made a gap-isolated eigenstate of a sparse lattice Hamiltonian, and that this holds in full-wave simulations and in measurements on 3D-printed acoustic samples.

The key result is the boundary law. The number of long-range couplings is set by the SC sites on the rim of the pattern, not by its  interior or by the size of the lattice, so a pattern whose amplitudes vary in a complicated way needs no more long-range couplings than a uniform pattern with the same support.
The law also helps in choosing targets, because outlines without SC sites need only nearest-neighbor couplings. This matters for platforms in which long-range couplings are hard to make, such as photonic crystals and coupled-resonator arrays.

The construction is not tied to the square lattice. 
It needs only a lattice graph with tunable couplings and on-site energies, so it should apply to other lattices and, under suitable conditions, to three-dimensional ones. 
There the SC sites lie on a surface, and the number of long-range couplings should scale with the surface area of the pattern, not with its volume.
A natural next step is to prescribe several eigenstates in one structure, which is possible only if their local interference conditions are compatible. 
Such structures could have designed degeneracies or orbital-like multiplets.

Because the target is the only mode near its frequency, it can be excited selectively, as in our measurements. 
In acoustic resonators, photonic structures or circuits\cite{lu2009phononic,zhang2018topologicalsound,ma2019topophases,lee2018topolectrical} , this could be used to deliver wave energy to chosen sites while suppressing the response elsewhere.
Experimental realization of complex-valued eigenstates would require control over coupling phases\cite{fang2012magneticfield,rechtsman2013floquet}, while nonlinear and non-Hermitian implementations would introduce additional constraints associated with nonlinear responses and non-Hermitian spectra\cite{xie2024parent,kirsch2021nonlinear}. 
In this approach the eigenstate, not the structure, is the starting point of the design.

	\paragraph{References}
	\bibliographystyle{naturemag_noURL}
	\bibliography{main}

@article{anderson1958,
  title   = {Absence of diffusion in certain random lattices},
  author  = {Anderson, P. W.},
  journal = {Phys. Rev.},
  volume  = {109},
  number  = {5},
  pages   = {1492--1505},
  year    = {1958},
  publisher = {American Physical Society},
  doi     = {10.1103/PhysRev.109.1492}
}

@article{yablonovitch1991donor,
  title   = {Donor and acceptor modes in photonic band structure},
  author  = {Yablonovitch, E. and Gmitter, T. J. and Meade, R. D. and Rappe, A. M. and Brommer, K. D. and Joannopoulos, J. D.},
  journal = {Phys. Rev. Lett.},
  volume  = {67},
  number  = {24},
  pages   = {3380--3383},
  year    = {1991},
  publisher = {American Physical Society},
  doi     = {10.1103/PhysRevLett.67.3380}
}

@article{akahane2003highq,
  title   = {High-{Q} photonic nanocavity in a two-dimensional photonic crystal},
  author  = {Akahane, Y. and Asano, T. and Song, B.-S. and Noda, S.},
  journal = {Nature},
  volume  = {425},
  number  = {6961},
  pages   = {944--947},
  year    = {2003},
  publisher = {Nature Publishing Group},
  doi     = {10.1038/nature02063}
}

@article{hsu2016bic,
  title   = {Bound states in the continuum},
  author  = {Hsu, Chia Wei and Zhen, Bo and Stone, A. Douglas and Joannopoulos, John D. and Solja{\v{c}}i{\'c}, Marin},
  journal = {Nat. Rev. Mater.},
  volume  = {1},
  pages   = {16048},
  year    = {2016},
  publisher = {Nature Publishing Group},
  doi     = {10.1038/natrevmats.2016.48}
}

@article{hasan2010ti,
  title   = {Colloquium: Topological insulators},
  author  = {Hasan, M. Z. and Kane, C. L.},
  journal = {Rev. Mod. Phys.},
  volume  = {82},
  number  = {4},
  pages   = {3045--3067},
  year    = {2010},
  publisher = {American Physical Society},
  doi     = {10.1103/RevModPhys.82.3045}
}

@article{lu2014topophotonics,
  title   = {Topological photonics},
  author  = {Lu, Ling and Joannopoulos, John D. and Solja{\v{c}}i{\'c}, Marin},
  journal = {Nat. Photon.},
  volume  = {8},
  number  = {11},
  pages   = {821--829},
  year    = {2014},
  publisher = {Nature Publishing Group},
  doi     = {10.1038/nphoton.2014.248}
}

@article{leykam2018flatband,
  title   = {Artificial flat band systems: from lattice models to experiments},
  author  = {Leykam, Daniel and Andreanov, Alexei and Flach, Sergej},
  journal = {Adv. Phys. X},
  volume  = {3},
  number  = {1},
  pages   = {1473052},
  year    = {2018},
  publisher = {Taylor \& Francis},
  doi     = {10.1080/23746149.2018.1473052}
}

@article{yu2011metasurface,
  title   = {Light propagation with phase discontinuities: generalized laws of reflection and refraction},
  author  = {Yu, Nanfang and Genevet, Patrice and Kats, Mikhail A. and Aieta, Francesco and Tetienne, Jean-Philippe and Capasso, Federico and Gaburro, Zeno},
  journal = {Science},
  volume  = {334},
  number  = {6054},
  pages   = {333--337},
  year    = {2011},
  publisher = {American Association for the Advancement of Science},
  doi     = {10.1126/science.1210713}
}

@article{liu2000lrsm,
  title   = {Locally resonant sonic materials},
  author  = {Liu, Zhengyou and Zhang, Xixiang and Mao, Yiwei and Zhu, Y. Y. and Yang, Zhiyu and Chan, C. T. and Sheng, Ping},
  journal = {Science},
  volume  = {289},
  number  = {5485},
  pages   = {1734--1736},
  year    = {2000},
  publisher = {American Association for the Advancement of Science},
  doi     = {10.1126/science.289.5485.1734}
}

@article{ma2019topophases,
  title   = {Topological phases in acoustic and mechanical systems},
  author  = {Ma, Guancong and Xiao, Meng and Chan, C. T.},
  journal = {Nat. Rev. Phys.},
  volume  = {1},
  number  = {4},
  pages   = {281--294},
  year    = {2019},
  publisher = {Nature Publishing Group},
  doi     = {10.1038/s42254-019-0030-x}
}

@article{lu2009phononic,
  title   = {Phononic crystals and acoustic metamaterials},
  author  = {Lu, Ming-Hui and Feng, Liang and Chen, Yan-Feng},
  journal = {Mater. Today},
  volume  = {12},
  number  = {12},
  pages   = {34--42},
  year    = {2009},
  publisher = {Elsevier},
  doi     = {10.1016/S1369-7021(09)70315-3}
}

@article{xue2019kagome,
  title   = {Acoustic higher-order topological insulator on a kagome lattice},
  author  = {Xue, Haoran and Yang, Yahui and Gao, Fei and Chong, Yidong and Zhang, Baile},
  journal = {Nat. Mater.},
  volume  = {18},
  number  = {2},
  pages   = {108--112},
  year    = {2019},
  publisher = {Nature Publishing Group},
  doi     = {10.1038/s41563-018-0251-x}
}

@article{wang2023zclassified,
  title   = {Realization of a {$\mathbb{Z}$}-classified chiral-symmetric higher-order topological insulator in a coupling-inverted acoustic crystal},
  author  = {Wang, Dongyi and Deng, Yuanchen and Ji, Jun and Oudich, Mourad and Benalcazar, Wladimir A. and Ma, Guancong and Jing, Yun},
  journal = {Phys. Rev. Lett.},
  volume  = {131},
  number  = {15},
  pages   = {157201},
  year    = {2023},
  publisher = {American Physical Society},
  doi     = {10.1103/PhysRevLett.131.157201}
}

@article{yang2015topoacoustics,
  title   = {Topological acoustics},
  author  = {Yang, Zhaoju and Gao, Fei and Shi, Xihang and Lin, Xiao and Gao, Zhen and Chong, Yidong and Zhang, Baile},
  journal = {Phys. Rev. Lett.},
  volume  = {114},
  number  = {11},
  pages   = {114301},
  year    = {2015},
  publisher = {American Physical Society},
  doi     = {10.1103/PhysRevLett.114.114301}
}

@article{ni2019hoti,
  title   = {Observation of higher-order topological acoustic states protected by generalized chiral symmetry},
  author  = {Ni, Xiang and Weiner, Matthew and Al{\`u}, Andrea and Khanikaev, Alexander B.},
  journal = {Nat. Mater.},
  volume  = {18},
  number  = {2},
  pages   = {113--120},
  year    = {2019},
  publisher = {Nature Publishing Group},
  doi     = {10.1038/s41563-018-0252-9}
}

@article{susstrunk2015mechanical,
  title   = {Observation of phononic helical edge states in a mechanical topological insulator},
  author  = {S{\"u}sstrunk, Roman and Huber, Sebastian D.},
  journal = {Science},
  volume  = {349},
  number  = {6243},
  pages   = {47--50},
  year    = {2015},
  publisher = {American Association for the Advancement of Science},
  doi     = {10.1126/science.aab0239}
}

@article{zhang2018topologicalsound,
  title   = {Topological sound},
  author  = {Zhang, Xiujuan and Xiao, Meng and Cheng, Ying and Lu, Ming-Hui and Christensen, Johan},
  journal = {Commun. Phys.},
  volume  = {1},
  pages   = {97},
  year    = {2018},
  publisher = {Nature Publishing Group},
  doi     = {10.1038/s42005-018-0094-4}
}

@article{schwartz2007,
  title   = {Transport and {Anderson} localization in disordered two-dimensional photonic lattices},
  author  = {Schwartz, Tal and Bartal, Guy and Fishman, Shmuel and Segev, Mordechai},
  journal = {Nature},
  volume  = {446},
  number  = {7131},
  pages   = {52--55},
  year    = {2007},
  publisher = {Nature Publishing Group},
  doi     = {10.1038/nature05623}
}

@article{painter1999,
  title   = {Two-dimensional photonic band-gap defect mode laser},
  author  = {Painter, O. and Lee, R. K. and Scherer, A. and Yariv, A. and O'Brien, J. D. and Dapkus, P. D. and Kim, I.},
  journal = {Science},
  volume  = {284},
  number  = {5421},
  pages   = {1819--1821},
  year    = {1999},
  publisher = {American Association for the Advancement of Science},
  doi     = {10.1126/science.284.5421.1819}
}

@article{hsu2013trapped,
  title   = {Observation of trapped light within the radiation continuum},
  author  = {Hsu, Chia Wei and Zhen, Bo and Lee, Jeongwon and Chua, Song-Liang and Johnson, Steven G. and Joannopoulos, John D. and Solja{\v{c}}i{\'c}, Marin},
  journal = {Nature},
  volume  = {499},
  number  = {7457},
  pages   = {188--191},
  year    = {2013},
  publisher = {Nature Publishing Group},
  doi     = {10.1038/nature12289}
}

@article{kazemi2023drawing,
  title={Drawing dispersion curves: band structure customization via nonlocal phononic crystals},
  author={Kazemi, Arash and Deshmukh, Kshiteej J and Chen, Fei and Liu, Yunya and Deng, Bolei and Fu, Henry Chien and Wang, Pai},
  journal={Phys. Rev. Lett.},
  volume={131},
  number={17},
  pages={176101},
  year={2023},
  publisher={APS}
}

@article{lucas2023tailoring,
  title={Tailoring microcombs with inverse-designed, meta-dispersion microresonators},
  author={Lucas, Erwan and Yu, Su-Peng and Briles, Travis C and Carlson, David R and Papp, Scott B},
  journal={Nat. Photon.},
  volume={17},
  number={11},
  pages={943--950},
  year={2023},
  publisher={Nature Publishing Group UK London}
}

@article{li2017controlling,
  title={Controlling propagation and coupling of waveguide modes using phase-gradient metasurfaces},
  author={Li, Zhaoyi and Kim, Myoung-Hwan and Wang, Cheng and Han, Zhaohong and Shrestha, Sajan and Overvig, Adam Christopher and Lu, Ming and Stein, Aaron and Agarwal, Anuradha Murthy and Lon{\v{c}}ar, Marko and others},
  journal={Nat. Nanotechnol.},
  volume={12},
  number={7},
  pages={675--683},
  year={2017},
  publisher={Nature Publishing Group UK London}
}

@article{melde2016holograms,
  title={Holograms for acoustics},
  author={Melde, Kai and Mark, Andrew G and Qiu, Tian and Fischer, Peer},
  journal={Nature},
  volume={537},
  number={7621},
  pages={518--522},
  year={2016},
  publisher={Nature Publishing Group UK London}
}

@article{kiczynski2022engineering,
  title={Engineering topological states in atom-based semiconductor quantum dots},
  author={Kiczynski, Mitchell and Gorman, Samuel K and Geng, Helen and Donnelly, Matthew B and Chung, Yousun and He, Yu and Keizer, Joris G and Simmons, Michelle Y},
  journal={Nature},
  volume={606},
  number={7915},
  pages={694--699},
  year={2022},
  publisher={Nature Publishing Group UK London}
}

@article{hafezi2013imaging,
  title={Imaging topological edge states in silicon photonics},
  author={Hafezi, Mohammad and Mittal, Sunil and Fan, J and Migdall, A and Taylor, JM},
  journal={Nat. Photon.},
  volume={7},
  number={12},
  pages={1001--1005},
  year={2013},
  publisher={Nature Publishing Group UK London}
}

@article{qi2019determining,
  title={Determining a local {Hamiltonian} from a single eigenstate},
  author={Qi, Xiao-Liang and Ranard, Daniel},
  journal={Quantum},
  volume={3},
  pages={159},
  year={2019},
  publisher={Verein zur F{\"o}rderung des Open Access Publizierens in den Quantenwissenschaften}
}

@article{chertkov2018computational,
  title={Computational inverse method for constructing spaces of quantum models from wave functions},
  author={Chertkov, Eli and Clark, Bryan K},
  journal={Phys. Rev. X},
  volume={8},
  number={3},
  pages={031029},
  year={2018},
  publisher={APS}
}

@article{lee2018topolectrical,
  title={Topolectrical circuits},
  author={Lee, Ching Hua and Imhof, Stefan and Berger, Christian and Bayer, Florian and Brehm, Johannes and Molenkamp, Laurens W. and Kiessling, Tobias and Thomale, Ronny},
  journal={Commun. Phys.},
  volume={1},
  pages={39},
  year={2018},
  doi={10.1038/s42005-018-0035-2}
}

@article{fang2012magneticfield,
  title={Realizing effective magnetic field for photons by controlling the phase of dynamic modulation},
  author={Fang, Kejie and Yu, Zongfu and Fan, Shanhui},
  journal={Nat. Photon.},
  volume={6},
  pages={782--787},
  year={2012},
  doi={10.1038/nphoton.2012.236}
}

@article{rechtsman2013floquet,
  title={Photonic {Floquet} topological insulators},
  author={Rechtsman, Mikael C. and Zeuner, Julia M. and Plotnik, Yonatan and Lumer, Yaakov and Podolsky, Daniel and Dreisow, Felix and Nolte, Stefan and Segev, Mordechai and Szameit, Alexander},
  journal={Nature},
  volume={496},
  pages={196--200},
  year={2013},
  doi={10.1038/nature12066}
}

@article{xie2024parent,
  title={Determining the non-{Hermitian} parent {Hamiltonian} from a single eigenstate},
  author={Xie, Xu-Dan and Xue, Zheng-Yuan and Zhang, Dan-Bo},
  journal={Phys. Rev. B},
  volume={110},
  pages={235113},
  year={2024},
  doi={10.1103/PhysRevB.110.235113}
}

@article{kirsch2021nonlinear,
  title={Nonlinear second-order photonic topological insulators},
  author={Kirsch, Marco S. and Zhang, Yiqi and Kremer, Mark and Maczewsky, Lukas J. and Ivanov, Sergey K. and Kartashov, Yaroslav V. and Torner, Lluis and Bauer, Dieter and Szameit, Alexander and Heinrich, Matthias},
  journal={Nat. Phys.},
  volume={17},
  pages={995--1000},
  year={2021},
  doi={10.1038/s41567-021-01275-3}
}

@book{Chu2005,
  author    = {Chu, Moody T. and Golub, Gene H.},
  title     = {Inverse Eigenvalue Problems: Theory, Algorithms, and Applications},
  publisher = {Oxford University Press},
  year      = {2005}
}

@article{Maimaiti2021,
  author  = {Maimaiti, Wulayimu and Andreanov, Alexei and Flach, Sergej},
  title   = {Flat-band generator in two dimensions},
  journal = {Phys. Rev. B},
  volume  = {103},
  pages   = {165116},
  year    = {2021},
  doi     = {10.1103/PhysRevB.103.165116}
}

@article{Sha2025,
  author  = {Sha, Yi-Xin and Xia, Ming-Yao and Lu, Ling and Yang, Yi},
  title   = {Efficient algorithms for the surface density of states in topological photonic and acoustic systems},
  journal = {Nat. Comput. Sci.},
  volume  = {5},
  pages   = {1192--1201},
  year    = {2025},
  doi     = {10.1038/s43588-025-00898-3}
}
	\newpage
	\clearpage
	
     \newpage
	\clearpage
	
	\newpage
	\clearpage
	
%
	\newpage
	\clearpage

	\newpage
	\clearpage
	
	\newpage
	\clearpage
	
\begin{methods}

\subsection{Acoustic implementation}
Each lattice site was represented by a vertical air-filled cavity of height $L_0=75~\mathrm{mm}$. 
Its fundamental longitudinal resonance, $f_0=c_0/(2L_0)\simeq2{,}287~\mathrm{Hz}$ for $c_0=343~\mathrm{m\,s^{-1}}$, served as the on-site resonance. 
The off-diagonal elements of each of the four Hamiltonians were quantized in increments of $0.1$.
All quantized nonzero elements were mapped to the calibrated tube geometries, with long-range terms implemented by directly connecting the corresponding spatially separated cavities.

The relation between channel geometry and effective coupling was calibrated using two identical coupled cavities. 
From their even- and odd-mode frequencies, $f_{\mathrm e}$ and $f_{\mathrm o}$, we defined $\Delta f=(f_{\mathrm e}-f_{\mathrm o})/(f_0)$ and $\bar f=(f_{\mathrm e}+f_{\mathrm o})/(2f_0)$, signifying the coupling strength and averaged frequency, respectively. 
Scanning the channel attachment height $Z_c$ and length $L_c$ produced a map from $(Z_c,L_c)$ to $(\Delta f,\bar f)$. 
Geometries satisfying $\bar f=1$ were selected to maintain the spectral center. 
The largest absolute off-diagonal element of each target Hamiltonian was assigned the largest calibrated coupling, and every other channel was selected to satisfy
\begin{equation}
\frac{t_{ij}}{t_{\mathrm{ref}}}
=\frac{H_{ij}}{\max_{m\neq n}|H_{mn}|},
\end{equation}
where $t_{ij}$ and $t_{\mathrm{ref}}$ denote the acoustic coupling between sites $i$ and $j$, and the reference acoustic coupling, respectively. 
The corresponding attachment height and channel length were then selected from the two-cavity calibration map to reproduce the required magnitude and sign.

\subsection{Full-wave simulations}
Air was assigned a density of $1.21~\mathrm{kg\,m^{-3}}$ and a sound speed of $343~\mathrm{m\,s^{-1}}$. 
All cavity and channel walls were treated as acoustically rigid. The complete cavity--channel structures were simulated so that the continuum calculation included finite cavity dimensions, the pressure variation in the channels, frequency-dependent coupling and residual interactions absent from the tight-binding model. 
Complex pressure fields were extracted directly from the eigenstates. 
Site-resolved fields were sampled on a horizontal plane at $z=73~\mathrm{mm}$, with their complex amplitudes retained for comparison with the prescribed real-valued profiles.

\subsection{Sample fabrication}
The four $12\times12$ cavity arrays were fabricated by stereolithographic three-dimensional printing using a photosensitive resin with a density of $1.15~\mathrm{g\,cm^{-3}}$ and a Young's modulus of $2{,}690~\mathrm{MPa}$. 
The internal air domains used the same geometric parameters as the full-wave models. 
The wall thickness was $1.5~\mathrm{mm}$. 
The large acoustic-impedance contrast between the resin and air allowed the solid walls to be approximated as acoustically rigid over the measured frequency range. 
Each cavity contained upper and lower circular access holes of radius $1.75~\mathrm{mm}$ for excitation and pressure detection.

\subsection{Acoustic measurements and field reconstruction}

A loudspeaker coupled to the lower access hole of a selected cavity supplied a signal with a linear frequency sweep. 
Signals from a reference probe in the upper access hole of the source cavity and a scanning probe inserted to the same depth in the upper access hole of each of the other cavities were recorded simultaneously using a Br\"uel \& Kj\ae r 3160-A-042 acquisition module. 
The source-cavity reference signal was also used as the local pressure measurement at that cavity.

For a measurement with the source placed at site $s_a$, let $S_i^{(a)}(f)$ and $S_{\mathrm{ref}}^{(a)}(f)$ denote the complex signals recorded by the scanning probe at site $i$ and by the source-cavity reference probe, respectively. 
The source-referenced complex response was defined as
\begin{equation}
p_i^{(a)}(f)
=
\frac{S_i^{(a)}(f)}
     {S_{\mathrm{ref}}^{(a)}(f)}
=
R_i^{(a)}(f)+\mathrm{i}I_i^{(a)}(f),
\end{equation}
where $R_i^{(a)}(f)$ and $I_i^{(a)}(f)$ are the real and imaginary parts of the response. 
This simultaneous complex referencing removes the excitation amplitude and phase common to each scan and maintains a consistent phase convention across the measured array.
The same reference probe, acquisition channels and electronic signal chain were retained throughout all measurements.

For the compact $S$, $P$ and $D$ states, a single source within the characteristic modal region was used. 
For the spatially extended CUHK-shaped state, eight independent scans were performed using two source positions in each letter. 
All eight source positions were selected at non-nodal sites belonging to the same-phase lobe of the target state. 
At the target frequency $f_t$, each source-referenced field was normalized according to
\begin{equation}
\widetilde{p}_i^{(a)}(f_t)
=
\frac{p_i^{(a)}(f_t)}
     {\displaystyle\max_j\left|p_j^{(a)}(f_t)\right|},
\qquad a=1,\ldots,8.
\end{equation}
The reconstructed CUHK field was then obtained by coherently averaging the eight normalized complex-response maps,
\begin{equation}
p_i^{\mathrm{CUHK}}(f_t)
=
\frac{1}{8}
\sum_{a=1}^{8}
\widetilde{p}_i^{(a)}(f_t).
\end{equation}

Because all source positions have the same relative modal phase and the response near $f_t$ is dominated by the same target resonance, the eight source-referenced fields share a common global phase convention. 
No additional source-dependent phase rotation was therefore applied before coherent averaging. 
For visualization, the coherently averaged field was normalized once more by its maximum absolute amplitude, and its real part was plotted using the common experimental reference phase.

%

\end{methods}

\paragraph{Data availability}
The dataset containing the Hamiltonian matrices and target-state definitions for the 116 case studies presented in the Supplementary Information, together with the accompanying case figures, is available at Zenodo: \url{https://doi.org/10.5281/zenodo.23023466}. The data underlying the figures in the main text are available from the corresponding author.

\paragraph{Code availability}
The \textsc{Shape} software (version 1.0.0), including the Windows x64 standalone application and Python source code, is available on Zenodo at \url{https://doi.org/10.5281/zenodo.23034833} under the MIT License. The included scripts regenerate the datasets and accompanying case figures for the cases presented in the Supplementary Information.

\paragraph{Acknowledgements}
B. X. acknowledges funding from the National Natural Science Foundation of China (Grants No. 62475225, No. 12404187), the National Key R\&D Program of China (Grant No. 2023YFA1407700, 2025YFA1412300), Guangdong Basic and Applied Basic Research Foundation (Grant No. 2024A1515012031),  Shenzhen Science and Technology Innovation Commission (Grant No. JCYJ20240813113619025), Guangdong Provincial Quantum Science Strategic Initiative (GDZX2604007).
\paragraph{Competing interests}
The authors declare no competing interests.
	
	\newpage
	\clearpage
	
\end{document}